\input harvmac
\input epsf
%\draftmode

\def\R{\relax{\rm I\kern-.18em R}}
\font\cmss=cmss10 \font\cmsss=cmss10 at 7pt
\def\Z{\relax\ifmmode\mathchoice
{\hbox{\cmss Z\kern-.4em Z}}{\hbox{\cmss Z\kern-.4em Z}}
{\lower.9pt\hbox{\cmsss Z\kern-.4em Z}}
{\lower1.2pt\hbox{\cmsss Z\kern-.4em Z}}\else{\cmss Z\kern-.4em
Z}\fi}\
\def\np{Nucl. Phys. }
\def\pl{Phys. Lett. }
\def\pr{Phys. Rev. }
\def\lmp{Lett. Math. Phys. }

\def\cmp{Comm. Math. Phys. }
\def\mpl{Mod. Phys. Lett. }
\def\p{\partial}

\def\CF{{\cal F}}
\def\CO{{\cal O}}

\def\CZ{{\cal Z}}

\def\l{\ell}

\def\CD{{\cal D}}

\def\CO{{\cal O}}
\def\Tr{{\rm Tr}}
\def\b{\beta}
\def\f{\Phi}
\def\char{\chi_{_h}}

\def\sumh{\sum_{h}}
\def\CN{{\cal N}}

\def\Oh{\Omega_{h}}
\def\ov{\over}
\def\dag{\scriptscriptstyle \dagger}
\def\barint{-\hskip -11pt\int}

\rightline{CERN-TH/96-311}
\rightline{hep-th/9611011}

\Title{}
{\vbox{\centerline{ Two-Dimensional Chiral Matrix Models}
\centerline{and   String Theories}   }}

\vskip6pt

\centerline{Ivan K. Kostov \footnote{$ ^\ast $}{member of
CNRS}\footnote{$ ^\dagger$}{{\tt kostov@spht.saclay.cea.fr}}}

\centerline{{\it C.E.A. - Saclay, Service de Physique Th\'eorique}}
 \centerline{{\it
  F-91191 Gif-Sur-Yvette, France}}

\bigskip

\centerline{Matthias Staudacher\footnote{$ ^\diamond$}{{\tt
matthias@nxth04.cern.ch}}}

\centerline{{\it CERN, Theory Division}}
 \centerline{{\it CH-1211 Geneva 23, Switzerland}}

\vskip .3in
\baselineskip10pt{
We formulate and solve a class of two-dimensional
matrix gauge models describing ensembles of non-folding
surfaces covering an oriented, discretized, two-dimensional
manifold. We interpret the models as string theories
characterized by a set of coupling constants associated to
worldsheet ramification points of various orders.
Our approach is closely related to, but simpler than, the
string theory describing two-dimensional Yang-Mills theory.
Using recently developed character expansion methods
we exactly solve the models for target space lattices of
arbitrary internal connectivity and topology.
  }
\bigskip

\rightline{SPhT-96/123}
\leftline{CERN-TH/96-311}
\leftline{November 1996}
%\draft
\Date{ }

\baselineskip=16pt plus 2pt minus 2pt
\bigskip
%%%%%%%%%%%%%%%%%%%%%%%%%%%%%%%%%%%%%%%%%%%%%%%%%%%%%%%%
\lref\WG{D.~Weingarten, \pl B 90 (1980) 280.}
\lref\EGU{T.~Eguchi and H.~Kawai, \pl B 114 (1982) 247
and \pl B 110 (1982) 143. }
\lref\DFJ{B.~Durhuus, J.~Fr\"ohlich and T.~J\'onsson,
\np B 240 FS[12] (84) 453.}
\lref\KSWI{V.A.~Kazakov, M.~Staudacher and T.~Wynter,
\cmp 177 (1996) 451.}
\lref\AFOQ{H. Awada, M. Fukuma, S. Odake and Y.-H. Quano, \lmp 31
(1994) 289.}
\lref\KSWII{V.A.~Kazakov, M.~Staudacher and T.~Wynter,
\cmp 179 (1996) 235.}
\lref\KSWIII{V.A.~Kazakov, M.~Staudacher and T.~Wynter,
\np B 471 (1996) 309.}
\lref\KSWR{V.A.~Kazakov, M.~Staudacher and T.~Wynter,
{\it Advances in Large $N$ Group Theory
and the Solution of Two-Dimensional $R^2$ Gravity},
hep-th/9601153,
1995 Carg\`ese Proceedings.}
\lref\GAUGE{A.A. Migdal, Zh. Eksp. Teor. Fiz. 69 (1975) 810
(Sov. Phys. JETP 42 (413)).}
\lref\DK{M.R.~Douglas and V.A.~Kazakov, \pl B 312 (1993) 219.}
\lref\RUS{B.~Rusakov, \mpl A5 (1990) 693.}
\lref\GT{D.~Gross, \np B B 400 (1993) 161;
D.~Gross and W.~Taylor, \np B 400 (1993) 181;
\np B 403 (1993) 395.}
\lref\MOO{S.~Cordes, G.~Moore and S.~Ramgoolam,
{\it Large N 2-D Yang-Mills Theory and Topological String Theory},
hep-th/9402107, and
{\it Lectures on 2D Yang-Mills Theory,
Equivariant Cohomology and Topological Field Theories},
hep-th/9411210, 1993 Les Houches and Trieste Proceedings;
G.~Moore, {\it 2-D Yang-Mills Theory and Topological Field Theory},
hep-th/9409044.}
%%%\lref \DK {M.R.~Douglas and V.A.~Kazakov, \pl B 319 (1993) 219.}
\lref\GW{D.J.~Gross and E.~Witten, \pr D 21 (1980) 446.}
 %\lref\SW{M.~Staudacher and T.~Wynter, in preparation.}
\lref\IKK{V. Kazakov, \pl B 128 (1983); K. O'Brien and J.-B. Zuber,
\pl B 144 (1984) 407; I.~Kostov, \np B 265 (1986), 223, B 415 (1994)
29.}
\lref\DOUG{M.R.~Douglas,
{\it Conformal Field Theory Techniques in
Large $N$ Yang-Mills Theory}, hep-th/9311130, 1993 Carg\`ese
Proceedings.}
\lref\RUDD{R.~Rudd, {\it The String Partition Function for
QCD on the Torus}, hep-th/9407176.}
\lref\DIJKI{R.~Dijkgraaf, {\it Mirror Symmetry and Elliptic Curves},
in
{\it The Moduli Space of Curves}, Progress in Mathematics 129
(Birkh\"auser, 1995), 149.}
\lref\DIJKII{R.~Dijkgraaf, {\it Chiral Deformations of Conformal
Field Theories}, hep-th/9609022.}
\lref\ITDI{P.~Di~Francesco and C.~Itzykson, Ann. Inst. Henri.
Poincar\'e Vol. 59, no. 2 (1993) 117.}
\lref\KSWIV{I.~Kostov, M.~Staudacher and T.~Wynter,
in preparation.}
\lref\KK{V.A.~Kazakov and I.~Kostov,
\np B 176 (1980) 199; V.A.~Kazakov, \np B 179 (1981) 283.}

%%%%%%%%%%%%%%%%%%%%%%%%%%%%%%%%%%%%%%%%%%%%%%%%%%%%%%
\newsec{Introduction}

The idea that the strong interactions are
described by a string theory which is in some sense dual to
perturbative QCD is a major challenge for
high energy theory.
 More generally, a  $D$-dimensional confining Yang-Mills theory is
expected to define a string theory with $D$-dimensional
target space stable in the interval $2\le D\le 4$.

This ``YM string'' has been constructed so far only
  for $D=2$ \GT,\MOO.
It has been shown that  the partition function
%$\CZ(A,G,N)$
of  pure  $U(N)$ gauge theory defined on
a two-dimensional manifold of given genus and  area
%$G$ and area $A$
can be represented in terms of a weighted sum over maps from a
worldsheet $\Sigma_W$ to spacetime $\Sigma_T$.
 The allowed  worldsheet configurations  represent {\it minimal area
maps} $\Sigma_W \rightarrow \Sigma_T$.
 The latter condition is equivalent to the condition that the
embedded surfaces are not allowed to have {\it folds}.
 Ramification points are however allowed, as it has been suggested in
earlier studies \KK .  In other words, the  path integral of the
``${\rm YM}_2$ string'' is over all  {\it branched covers}
of the target manifold \GT , \MOO.

Unfortunately, this construction is highly involved and
does not easily
reduce to a system of simple geometrical
principles\foot{The
random surface representation of    $YM_2$ on a lattice found in
\IKK\ is geometrically clear  but has the inconvenience of being
highly redundant. For example, the folds are not forbidden but their
contribution vanishes as a result of cancellations.}.
One can nevertheless speculate, using the analogy with the
random-walk representation of the $O(N)$ model,
that there exists an underlying $2D$ string theory
with clear geometrical interpretation.
 The ${\rm YM}_2$ string is obtained from the latter  by tuning the
interactions
due to ramification points and adding new contact interactions via
microscopic tubes, etc.

One is thus led to look for the most  general $2D$ string theory
whose
path integral is given by branched covers of the target space.  Any
such string theory is invariant under area-preserving
diffeomorphisms of the target space $\Sigma_T$. Its partition
function depends on $\Sigma_T$ only through its  genus
and  area.
In addition to the topological  coupling constant $N^{-1}$ there is a
set of couplings $t_n$    associated with the  interactions due to
ramification points of order $n$.

 In this letter we propose a discretization
of the path integral for such  string theories,
 in which the target space  $\Sigma_T$ represents a
two-dimensional simplicial complex with given topology.   The branch
points (the images of the ramification points) are thus located at
the vertices of the target lattice  $\Sigma_T$.
We will not try to establish
a worldsheet description but instead  construct and solve a class of
equivalent matrix models.  Our approach is thus
analogous to quantizing Polyakov string theory by employing
randomly triangulated surfaces.

The  matrix model  associated with the target space $\Sigma_T$ will
be formulated as a $2D$ lattice gauge theory whose  local link
variables are complex $N\times N$ matrices.
 However, the matrices are no
longer unitary:  the Haar measure on $U(N)$  is replaced by a
Gaussian measure.  The model  resembles very much the $2D$ Weingarten
model \WG\ with the
important difference that the lattice action  represents a sum
over the {\it positively} oriented cells only.   The latter
restriction  eliminates from the string path integral all surfaces
containing
folds (which are believed to cause the trivial critical behavior of
the  standard Weingarten model \DFJ   ).
We will demonstrate
that these  $2D$ matrix models  are  exactly solvable for any
target space lattice and any $N$ by employing the character
expansion methods recently developed in \KSWI, \KSWII, \KSWIII,
\KSWR.
We pay special attention to the cases of spherical and toroidal
topology. In the first case we observe that
the string theory exists only for a target space with sufficiently
large area.
At the critical area a third-order transition takes place due to the
entropy of the ramification points.
In the second case we find
the same partition function  as the chirally perturbed  conformal
field theory considered recently by R.~Dijkgraaf \DIJKII.
 In this letter we restrict our attention to the discrete case since
it involves combinatorial problems interesting on their own.
The continuum limit  of
an infinitely dense target lattice  will be
 considered in detail elsewhere \KSWIV.

 \newsec{Definition of the models}

 By target space  $\Sigma_T$  we will understand  an oriented
triangulated surface
(two-dimensional simplicial complex) of genus $G$
  containing $\CN_0$ points, $\CN_1$ links and
$\CN_2$ two-dimensional cells.
These numbers are related by the Euler formula
\eqn\eiL{ \CN_0-\CN_{1} +\CN_2 = 2-2G.}
In addition, each cell $c$ is characterized by its
area $A_c$.

We will first consider the simplest string theory with this  target
space, in which all ramification points have Boltzmann weight one,
and the corresponding  matrix
model.  The generalized model,
to be considered in sect.~6,
will depend on a set of external field variables associated with the
points of  $\Sigma_T$.

At each link $\l$
is defined a field variable $\Phi_\l$ representing an $N\times N$
matrix  with complex elements.
The  partition function of the matrix model is defined as
 \eqn\cwm{
\CZ=\int \prod_{\l} [\CD \f_{\l}] \prod_c
{}~\exp (\beta_c~N~\Tr   \f_{c} ), }
where $\beta_c = e^{-A_c}$ , $\Phi_c$     denotes the ordered product
$\prod_{\l\in \p c}\Phi_\l $ of link variables along the oriented
boundary $\p c$ of the cell $c$,
 and the integration over the link variables is performed with the
Gaussian measure
\eqn\measure{
 [\CD \Phi_{\l}]=({N /\pi})^{N^2}
\Pi_{ij} d\Phi_{ij} d\Phi_{ij}^{*}
 e^{-N~\Tr~\Phi_{\l}^{ } \f_{\l}^{\dag} }.}
  Note that we omit the complex conjugate $\f_{c}^{\dag}$ from the
plaquette action,
eliminating thereby orientation reversing plaquettes.  This is why we
 call the model
with partition function \cwm\ a {\it chiral} matrix model. The  model
is invariant under  complex conjugation of the matrix variables  {\it
and}  reversing   the orientation  of the target space.

The perturbative expansion of \cwm\  results in a
representation of the free energy $\CF=\ln \CZ$ in terms
of connected  lattice   surfaces $\Sigma_W$ embedded in the target
surface $\Sigma_T$.
 Each of these surfaces is obtained by placing
plaquettes on the faces of the lattice $\Sigma_T$ and gluing any
two together along the edges.
All plaquettes should  have the same orientation, which  means that
the  surfaces  cannot have folds.
It is geometrically evident that the number of plaquettes, say  $n$,
covering each cell is constant throughout
$\Sigma_T$. Thus the surface  $\Sigma_W$ is
wrapping $n$ times $\Sigma_T$ and its area is
$A_W= n A_T$ where
$A_T =\sum _c A_c$ is the total area of the
target space.  The surface $\Sigma_W$ may have
ramification points  whose images  are
points of $\Sigma_T$. The map $\Sigma_W\rightarrow \Sigma_T$ defines
at each point $p\in \Sigma_T$
a branching number $B_p$. The branch points
are the points with
 $B_p\ne 0$.
By the Riemann-Hurwitz formula
$2g-2 = n(2G-2)+\sum_p B_p$
where $g$ is the genus of the surface $\Sigma_W$.
 The free energy $\CF=\ln \CZ$ of our  chiral
model can be written
as
\eqn\random{
\CF=  \sum_{n=1}^{\infty}e^{-nA_T}
\sum_{g=G}^{\infty}N^{2-2g}  \CF(n, g| G, \CN_0)}
where $\CF(n, g |G, \CN_0 )$ is the number
of  no-fold surfaces of genus $g$ covering  $n$  times  $\Sigma_T$.
This number depends on $\Sigma_T$ only through its genus $G$ and the
number of the $\CN_0$ of
allowed locations of the branched points.

\newsec{Exact solution by the character expansion method}

 Applying to \cwm\ the same strategy as in ref.\KSWR , we expand the
exponential of the action for each  cell $c$
as a sum over the  characters
$\char $ of the polynomial representations of $GL(N)$. We use the
shifted weights
$h=\{h_1,h_2,\ldots,h_N\}$ where $h_i$ are related to the lengths
$m_1,..., m_N$ of the rows of the Young tableau by $h_i=N-i+m_i$ and
are therefore subjected to the constraint $h_1>h_2>\ldots>h_N\geq 0$.
 We will denote by  $|h|=\Sigma_i m_i$  the total number of boxes of
the Young tableau and by
\eqn\dim{\Delta_h=\prod_{i<j} { h_i-h_j \ov  i-j}.}
the dimension of the representation $h$.
 The  character expansion of the exponential  then   reads
\eqn\charexp{
e^{\b_c N\Tr \f_c} = \sum_{ h } \b_c^{|h|}
{}~{\Delta_h \ov \Omega_h}~\char(\f_c),}
where we have defined
\eqn\omega{\Omega_h  = N^{-|h|}~\prod_{i=1}^{N}
{h_{i}! \ov (N-i)!}.}
%
%\eqn\aone{
%\chi_{_h}(A_1)=N^{|h|}
%~\prod_{i=1}^{N} {(N-i)! \ov h_{i}!}
%~\Delta_h.}
%

Now the model can be diagonalized in $h$-space  by using the
following   two simple but powerful facts.
Firstly, one easily proves \KSWI,\KSWR
\eqn\gamma{
\int [\CD \f]~\char(\f \f^{\dag}) =  \Omega_h~\Delta_h,}
where $\Omega_h$ is given as above by \omega.
 Secondly, one uses that complex matrices may be diagonalized
by a bi-unitary transformation $U \f_D V^{\dag}$, where $U,V$ are
unitary and $\f_D$ is a positive definite and diagonal matrix.
Then one uses the formula \gamma\ together with the
 fission and fusion rules for unitary matrices
\eqn\ff{\eqalign{
\int \CD U~\char(A U B U^{\dag})&= {1 \over
\Delta_h}~\char(A)~\char(B), \cr
\int \CD U~\char(A U)~\chi_{ _ {h'}}(U^{\dag} B) &=
\delta_{h, h'}~{1 \over \Delta_h}~\char(AB), \cr } }
to derive fission and fusion rules for complex matrices corresponding
to the measure \measure :
\eqn\cff{\eqalign{
\int [\CD \f]~\char(A \f B \f^{\dag})&=
{\Omega_h \over \Delta_h}
{}~\char(A)~\char(B), \cr
\int [\CD \f]~\char(A \f)~\chi_{_ {h'}}(\f^{\dag} B) & =
\delta_{h, h'}~{\Omega_h \over \Delta_h}~\char(AB). \cr }}
Therefore, complex matrices behave exactly like unitary matrices
apart from the extra local factor $\Omega_h$ for each
integration\foot{
The fission and fusion rules actually hold  for any $U(N)$ invariant
measure $[\CD\Phi]$, if we define the factor
$\Omega_h$ by the integral \gamma .
In particular one could consider potentials higher than
linear in $\f \f^{\dag}$ in \measure. }.

Having at hand \cff, we can proceed with the exact solution
of \cwm, just as in the gauge theory case \GAUGE,\RUS.
Using the fusion rule one progressively eliminates links
between adjoining cells until one is left with
a single  cell.
The remaining links along that plaquette are integrated
out by applying the fission rule.
One arrives at the formula
\eqn\solution{
\CZ=\sumh \Delta_h^{2 - 2 G}~\Omega_h
 ^{\CN_0 + 2 G -2}~
e^{-|h|A_T }.}
 We observe that, in accord  with the
surface representation \random ,  the solution's only dependence on
the original target
space lattice data is through the genus $G$, the
number of vertices $\CN_0$, and the total area
$A_T$ of the
target space.
It is therefore clear that a given
character expansion \solution\ corresponds to many different
%\foot{
%The fact that all these models have the same partition function
%does of course not mean that the matrix models are identical:
%Correlators will be different, in general.}
complex multi matrix models \cwm\ with the same
partition function: As many, as we can form simplicial
complexes differing in their internal connectivity but
constrained by the global data $G$, $\CN_0$ and $A_T$.   (This is a
lattice version of the
invariance with respect to area-preserving automorphisms.)
 In particular we can represent  the target space by a one-cell
complex and write down a matrix model
with a minimal number $\CN_1=\CN_0-1+2G$
of link variables $\f_{\l}$ giving the expansion \solution:
 \eqn\onePG{
 \CZ =\int \prod_{\ell=1}^{\CN_0-1+2G}
 [\CD \f_{\ell}] 
%\prod_{s=\CN_0}^{2 G} [\CD \f_{s}][\CD \f_{\tilde 2 s}]
%\prod_{k=1}^{\CN_{0}-1}  [\CD \f_{k}]
~\exp\Bigg(e^{-A_T}~N~ \Tr\Big[ \prod_{s=1}^G \f_{2 s-1}^{}\f_{2 s}^{}
\f_{2 s-1}^{\dag}
\f_{2 s}^{\dag}
\prod _{k= 2 G+1}^{\CN_{0}-1 + 2G}
\f_{k}^{} \f_{k}^{\dag}\Big] \Bigg) .}
%
%This is an effective one-plaquette model.
%for a target space of
%arbitrary genus $G$, area $A$ and $\CN_v$ vertices.
%It is still not at all unique:
%We can cut up the target space manifold in
%many different ways. Each possibility corresponds to  one of the
% %spanning trees of the target lattice.

Note that the trace in the exponent of \onePG\
corresponds to the element of the homotopy group of a surface with
$G$ handles and $N_0$ punctures which is equivalent to the identity
(Fig. 1).
   %
%%%%%%%%%%%%%%%%%%%%%%%%%%%%%%%%%%%%%%%%%%%%%%%%%%%%%%%%%%%
\vskip 20pt
\hskip 70pt
\epsfbox{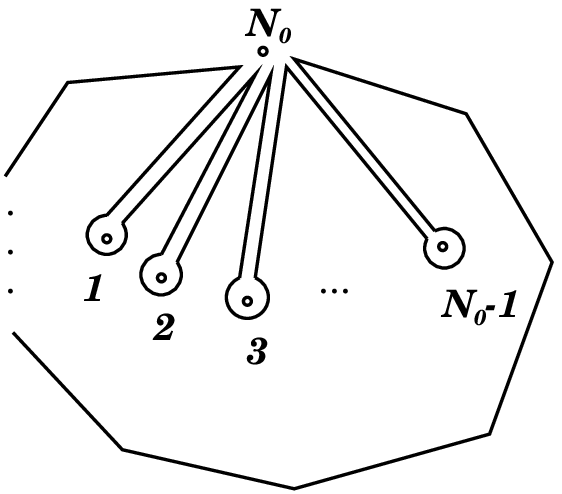}
\vskip 30pt
%%%%%%%%%%%%%%%%%%%%%%%%%%%%%%%%%%%%%%%%%%%%%%%%%%%%%%%%%%%%%%
\centerline{Fig.1: The target space of the effective one-plaquette model.}

\newsec{Spherical target space  }

 In the case of a spherical target space ($G=0$) the
partition function \solution\ becomes:
\eqn\sphere{
\CZ_{G=0}=\sumh \Delta_h^{2}~\Omega_h
 ^{\CN_0 -2}~
e^{-|h|A_T} .}
The simplest example of $\CN_0=2$ is described by a Gaussian
integral.
   There is only one surface covering $n$ times a sphere with two
punctures and
it contributes ${1 \ov n} e^{-nA_T} $ to the free energy
(the factor $1 \ov n$ coming from the cyclic symmetry). No
higher-genus surfaces are contributing. Summing on $n$ gives
therefore
\eqn\gzero{
\CF=-N^2\ln(1-e^{-A_T}),}
which  is also the result of the Gaussian integration.

If we have $\CN_0 \geq 3$ the situation becomes non-trivial.
Now there are contributions from non-spherical ($g > 0$)
world sheets as well. Concentrating on the leading
($g=0$, i.e.~$N=\infty$) term, we can proceed by
using the saddlepoint techniques of \KSWI.
Here the sum \sphere\ describes a ``gas'' of
mutually repulsive weights at thermal equilibrium and
is dominated by  the  contribution of a ``classical'' configuration
$\{ h \}$.
One introduces continuum variables
$h={h_i \over N}$ and a density $\rho(h)={1 \over N}\sum_i
\delta(h-h_i)$.
The saddlepoint density is found by functionally varying \sphere\
with respect to $h$; this gives
\eqn\saddle{
\barint_b^a dh' {\rho(h') \over h-h'}=
{2 -  \CN_0 \over 2}\log h +{1 \over 2} A_T  - \log {h \over h-b}. }
It is straightforward to explicitly solve this equation and
calculate the free energy. Here we will merely write out the
equation determining the critical behavior; one finds, for the
variable $u={1 \over 4} (\sqrt{a} + \sqrt{b})^2$, the
algebraic equation
\eqn\critical{ (\CN_0-1)~e^{-A_T}~u^{\CN_0-1} = u - 1  .}
This equation predicts for $\CN_0 \geq 3$ a critical
area $A_T=A_T^c >0$: The sum over surfaces diverges
if the area of the target manifold falls below
$A_T^c$.
 A quite similar  transition has already observed in $YM_2$ by
Douglas and Kazakov \DK .
One finds for the free energy near the critical
area
\eqn\pure{
\CF \sim N^2 ( A_T- A_T^c)^{5 \ov 2}  }
i.e., the same critical behavior as the Polyakov string in zero
dimensions.  We can understand the critical behavior by
 considering e.g.~$\CN_0=3$
\eqn\three{\CZ=\int [\CD \f_1] [\CD \f_2]
{}~e^{q N~ \Tr \f_{_1}  \f_1^{\dag}  \f_{_2} \f_2^{\dag} }.}
where $q=e^{-A_T}$. Eq.\three\ generates
planar diagrams which can
be interpreted as abstract plaquettes without embedding. Its
 critical behavior is in the university class of
 pure gravity.

 For $\CN_0 \geq 4$ the models
\onePG\ are intractable
without our character expansion techniques.
The qualitative behavior remains the same:
As seen from \critical, we can always find
the behavior of pure $2D$ gravity.
%They exists only if the area of the target manifold is sufficiently
% large.
The critical size of the target sphere is
given by
\eqn\crAr{A_T^c = 2\ln (\CN_0-1) +
(\CN_0-2)\ln \Big({\CN_0-1\over \CN_0-2}\Big) .}
 and tends to infinity  in the continuum limit  $\CN_0 \to\infty$.
It is therefore evident that   the theory
should be modified in order to have a sensible  continuum limit.

  \newsec{Toroidal target space}

If the target space is a torus ($G=1$) the Vandermonde determinants
in \solution\ cancel,
%and the partition function becomes
%
%\eqn\torsum{
%\CZ_{G=1}=\sumh \Omega_h
% ^{\CN_0 }~
%e^{-|h|A_T} .}
%
 the free energy is now order $\CO(N^0)$
and the
  saddlepoint method  is no longer
applicable.  However, we can do even better in this case and
immediately give
a result to all orders in $1 \ov N^2$.
Each term in the sum \solution\
%\torsum\
factorizes into a product over the rows
of the Young
  tableau; the sum remains nevertheless non-trivial
due to the ordering constraint on the weights.
Using \omega, rewrite $\CZ$ as
\eqn\content{
\CZ_{G=1}=\sumh e^{-|h|A_T} \prod_{\{i,j\}\in h}
\Big(1 - {i-j \over N}\Big)^{\CN_0}.}
Here $\{i,j\}$ denotes the a box in row $i$ and column
$j$ and the product goes over all boxes in the tableau $h$.
Slicing the tableau through the diagonal and counting
the fraction of boxes in the rows of the upper half
and the columns of the lower half we can elegantly
express \content\ as
 \eqn\theta{
\CZ_{G=1}=\oint {dz \over 2 \pi i z}\prod_{n=0}^{\infty}
\big[1 + z q^{n+\half} \prod_{k=1}^{n}
\Big(1+{k \over N}\Big)^{\CN_0} \big]
\big[1 + z^{-1} q^{n+\half} \prod_{k=1}^{n}
\Big(1-{k \over N}\Big)^{\CN_0} \big]}
where we have denoted $ q=e^{-A_T}$.
The contour integral ensures that an equal number of rows
and columns emanate from the diagonal; it eliminates
configurations that cannot be interpreted as a tableau\foot{It is
interesting to point out that a very similar result
was first found by Douglas \DOUG \ in the context of continuum $2D$
Yang-Mills theory on the torus.
The  combinatorial derivation of \theta\
 is due to  Dijkgraaf  \DIJKI.}.
 The leading  order of the free energy
gives the partition function of the noninteracting   string. It is
immediately
extracted from either \content\ or \theta\ and is identical to what
one finds for the
continuum Yang-Mills theory on a torus \GT:
\eqn\fretor{
\CF=
-\sum_{n=1}^{\infty} \ln(1-q^n) + \CO(N^{-2}).}
The differences start with the
  ${1 \ov N^2}$  terms.  These are in principle computable in terms
of quasi modular forms as in
\DOUG,\RUDD,\DIJKI,\DIJKII.

Similarly to the ${\rm YM}_2$ theory,  the matrix models with
toroidal target spaces  can be  interpreted as  chiral deformations
of a two-dimensional free field
theory defined by the action  \DOUG ,\DIJKI,\DIJKII
\eqn\acdp{S= \int{d^2z\over 2\pi} \p\varphi\bar\p \varphi+ \oint
\sum_n {s_n\over n+1} (\p\varphi)^{n +1}.}
The partition function \theta\ is equal to the zero $U(1)$ charge
sector of the
partition function of \acdp\ computed on a torus,
with an appropriate choice of the couplings $\{s_n\}$. %The
% Yang-Mills case corresponds to
%$s_n={1 \over N} \delta_{n,3}$ \DOUG.

 Let us mention an interesting interpretation of the model with
$\CN_0=1$ whose target space is a torus with one puncture.
  Explicitly, we have
\eqn\torus{\CZ=\int [\CD \f_{_1}] [\CD \f_{_2}]
{}~e^{q N~ \Tr \f_1^{} \f_2^{} \f_1^{\dag} \f_2^{\dag}}.}
 The model is superficially very similar to \three\ of the
last section. Again a network of square plaquettes (corresponding
to the vertices of \torus) is
generated. But here a local rule suppresses all {\it positive}
curvature:
%It generates
%the ensemble of all abstract discrete surfaces made of squares and
% having {\it %nonpositive} local curvature \IK.
The coordination numbers at the vertices of such a surface can take
values $4,8,12,...$, which correspond
to zero or negative curvature. No local curvature fluctuations,
leading to pure quantum gravity behavior, are possible: For fixed
genus
$g$ the surfaces contain only a finite number of quantized,
negative curvature defects of total curvature $-4 \pi (g-1)$.
The model is therefore similar
in spirit to the ``almost flat planar graphs'' of \KSWII, where
discrete surfaces of vanishing or positive local curvature
were considered.

\newsec{The general case}

 Let us now consider the discretized string theory in the general
case when  a ramification point of order $n =2,3,4,...$ associated
with the point $p\in \Sigma_T$ is weighted by
a factor  $t_n^{(p)}  $.
 It is straightforward to appropriately modify the original matrix
theory and solve it by repeating the same steps explained in
section 3.

   We will parametrize the couplings $t_m^{(p)}$ by
an  $N\times N$ external matrix field
   \eqn\cpL{t_n^{(p)}= {1\over N} \Tr (B_p)^n}
and insert the matrix $B_p$ into the
r.h.s. of \cwm .
This should be done in the following way: For
each point $p$ we choose one of the cells
$c$ such that $p\in\p c$
    and insert the matrix $B_p$ into the corresponding trace.
For example, the one-plaquette theory
\onePG\ becomes
 \eqn\genPG{
 \CZ =\int \prod_{\ell=1}^{\CN_0-1+2G}
 [\CD \f_{\ell}] 
%\prod_{s=\CN_0}^{2 G} [\CD \f_{s}][\CD \f_{\tilde 2 s}]
%\prod_{k=1}^{\CN_{0}-1}  [\CD \f_{k}]
~\exp\Bigg(e^{-A_T}~N~ \Tr\Big[ B_{\CN_0}
\prod_{s=1}^G \f_{2 s-1}^{}\f_{2 s}^{}
\f_{2 s-1}^{\dag}
\f_{2 s}^{\dag}
\prod _{k= 2 G+1}^{\CN_{0}-1 + 2G}
B_k \f_{k}^{} \f_{k}^{\dag}\Big] \Bigg) .}

 Then one can proceed
as in sect.~3, arriving at the following final expression replacing
\solution:
\eqn\gensol{
\CZ =\sumh \ e^{-|h|A_T}\  \bigg({\Delta_h\over\Oh}\bigg) ^{2-2 G} \
\prod_{k=1}^{\CN_0} \bigg(  \char(B_k )  {\Oh \over \Delta_h}
\bigg). }

The model \gensol\ can be solved in the spherical limit using the
methods developed in
 \KSWI-\KSWR.
Let us note that in the case of a sphere with three punctures $(G=0,
\CN_0=3)$ one
reproduces the
model of the dually weighted planar graphs in the most general
formulation presented in \KSWR:
\eqn\exdwg{\CZ_{(A_T=0,G=0,\CN_0=3 )}\ =
\sumh \ {\Oh \ov \Delta_h} \ \char(B_1 ) \char(B_2 )
\char(B_3 ). }

 \newsec{Concluding remarks}

1. We have constructed a discretization of the most general string
field theory of nonfolding surfaces  immersed in a two-dimensional
compact spacetime.
 In the continuum limit
the sum over the positions of a branch point should be  replaced by an
integral with respect to the area.  This means that the couplings
$t_n$  should   scale   as
\eqn\ctl{t_n = {A_T\over\CN_0} \tilde t_n}
where $\tilde t_n$ are the  interaction constants
of the continuum theory.   The properties of the expansion \gensol\ in this 
limit will be studied elsewhere \KSWIV .

2.  The YM-string should be obtained by a special tuning of the  coupling
constants and by introducing
contact (tube-like) interactions. These interactions can be
implemented by considering the $B$-matrices as dynamical
fields.

 3. We  point out an interesting  equivalence  between  ensembles
of  coverings of a sphere with fixed number of punctures
and an ensemble of abstract (i.e., without embedding)
 planar graphs. It is therefore not a miracle that we  find a third
order phase  transition as in the case of pure $2D$ gravity.   In our
case the transition is  due to the entropy of
surfaces with a large number of ramification points. A very similar
transition has already observed in $YM_2$ by Douglas and Kazakov \DK.

4.  The chiral matrix models  defined on a torus give a nice geometrical interpretation of the
general
chiral deformation of
topological theory  studied by R.~Dijkgraaf \DIJKII.
 The sum over surfaces can
again be interpreted in terms of abstract
 planar graphs describing
 random surfaces  with non-positive local curvature.

\vskip 30pt
\hskip -19pt{\bf Acknowledgements}

One of us (I.K.) thanks the CERN Theory Division for hospitality.

\listrefs

\bye